\documentclass[aps,prd,reprint,nofootinbib,longbibliography]{revtex4-1}

\usepackage{amsmath}
\usepackage{mathtools}
\usepackage{graphicx}
\usepackage[normalem]{ulem}
\usepackage[com pat=1.1.0]{tikz-feynman}
\usepackage{tikz}
\usepackage{aas_macros}
\usetikzlibrary{external}
\usepackage{comment}
\usetikzlibrary{arrows}
\usepackage[colorlinks=true, allcolors=blue]{hyperref}

\begin{document}

\title{New method for detecting fast neutrino flavor conversions in core-collapse supernova models with two-moment neutrino transport}

\author{Hiroki Nagakura}
\email{hirokin@astro.princeton.edu}
\affiliation{Department of Astrophysical Sciences, Princeton University, 4 Ivy Lane, Princeton, NJ 08544, USA}
\author{Lucas Johns}
\email[]{NASA Einstein Fellow \\ ljohns@berkeley.edu}
\affiliation{Department of Physics, University of California, Berkeley, CA 94720, USA}

\begin{abstract}
Fast-pairwise neutrino oscillations potentially affect many aspects of core-collapse supernova (CCSN): the explosion mechanism, neutrino signals, and nucleosynthesis in the ejecta. This particular mode of collective neutrino oscillations has a deep connection to the angular structure of neutrinos in momentum space; for instance, the appearance of electron neutrinos lepton number (ELN) angular crossings in momentum space is a good indicator of occurrences of the flavor conversions. However, many multi-dimensional (multi-D) CCSN simulations are carried out with approximate neutrino transport (such as two-moment methods), which limits the access to the angular distributions of neutrinos, i.e., inhibits ELN-crossing searches. In this paper, we develop a new method of ELN-crossing search in these CCSN simulations. The required data is the zero-th and first angular moments of neutrinos and matter profile, all of which are available in CCSN models with two-moment method. One of the novelties of our new method is to use a ray-tracing neutrino transport to determine ELNs in the direction of the stellar center. It is designed to compensate for shortcomings of the crossing searches only with the two angular moments. We assess the capability of the method by carrying out a detailed comparison to results of full Boltzmann neutrino transport in 1D and 2D CCSN models. We find that the ray-tracing neutrino transport improves the accuracy of crossing searches; indeed, the appearance/disappearance of the crossings is accurately detected even in the region of forward-peaked angular distributions. The new method is computationally cheap and has a benefit of efficient parallelization; hence, it will be useful for ELN-crossing searches in any CCSN models employed two-moment neutrino transport.
\end{abstract}
\maketitle

\section{Introduction}\label{sec:intro}
Neutrinos have been recognized as one of the key players in core-collapse supernova (CCSN). It has been suggested that the energy, momentum, and lepton number exchange between matter and neutrinos sways the destiny of the collapsing star; for instance, the neutrino heating in the post-shock flows revitalizes the stagnated-shock wave, i.e., triggering the onset of explosion (see recent reviews, e.g., \cite{2017hsn..book.1095J,2020LRCA....6....3M,2020LRCA....6....4M,2021Natur.589...29B}). The nuclear compositions in the ejecta is also dictated by neutrino-matter interactions, which sensitively depends on the energy spectrum of all flavor of neutrinos \cite{1996ApJ...471..331Q,1999PhRvL..82.5198H,2010ApJ...722..954R,2013ApJ...771...27Y,2014JPhG...41d4008M,2018JPhG...45a4001E,2018ApJ...852...40W,2019MNRAS.488L.114F,2020ApJ...904..163S,2021MNRAS.502.2319F,2021MNRAS.502.2319F}. From observational points of view, neutrino signals offer a diagnostic for the CCSN dynamics \cite{2013PhRvL.111l1104T,2013ApJS..205....2N,2016MNRAS.461.3296N,2019ARNPS..69..253M,2019arXiv191110656L,2020ApJ...898..139W,2021MNRAS.500..696N,2021arXiv210211283N,2021PTEP.2021a3E01S}. They are detectable for nearby CCSNe by currently-operating and future-planned terrestrial neutrino detectors (see recent reviews, e.g., \cite{2016NCimR..39....1M,2018JPhG...45d3002H}). Coincident neutrino detections by multiple observatories with different reaction channels will shed light on flavor-dependent features of neutrinos, which provides a precious information on deciphering CCSN dynamics and subsequent neutron star or black hole formation \cite{2011PhRvD..83k3006D,2019PhRvD..99l3009L,2021MNRAS.500..319N}. The accurate determination of neutrino radiation field is, hence, one of the fundamental tasks towards comprehensive understanding of death of massive stellar collapse.

A complete description of neutrino dynamics requires solving multi-dimensional (multi-D), multi-flavor, multi-energy, and multi-angle neutrino transport equations. Assuming that neutrino flavor conversions are suppressed by high matter density \cite{2000PhRvD..62c3007D}, the transport equation can be given by a Boltzmann equation. Recently, multi-D CCSN models with full Boltzmann neutrino transport have become available, which allows us to access the 3D features of neutrino momentum space \cite{2018ApJ...854..136N,2019ApJ...872..181H,2019ApJ...880L..28N,2020ApJ...903...82I}. On the other hand, it has been recognized that the assumption is not valid if instabilities of flavor conversion are turned on by neutrino-self interactions \cite{2010ARNPS..60..569D}, which is also known as collective neutrino oscillations. In this case, the transport equation needs to be altered from Boltzmann equations \cite{1993PhRvL..70.2363R,1999PhRvD..59l5011S,2000PhRvD..62i3026Y,2013PhRvD..87k3010V,2014PhRvD..89j5004V,2015PhLB..747...27C,2015IJMPE..2441009V,2016PhRvD..94c3009B}. The mean-field quantum kinetic equation seems to be the simplest extension but captures some essential features of flavor conversions. It involves, however, technical difficulties in the numerical treatments, since the self-interaction is highly nonlinear phenomena and there is a striking disparity of both spatial- and time scales between flavor conversions and CCSN dynamics. These practical issues have been obstackled for the detailed study of flavor conversions (but see recent efforts, e.g., \cite{2019PhLB..790..545A,2019PhRvD..99l3014R,2019PhRvD.100b3016M,2019PhRvL.122i1101C,2020PhRvD.101f3027S,2020arXiv201101948T,2020JCAP...06..048S,2021PhRvL.126f1302B,2021arXiv210101278M,2021PhRvD.103h3013R,2021arXiv210410532Z,2021JCAP...01..014S,2021arXiv210312743S}). For these reasons, the actual impact of flavor conversions on CCSN dynamics and the observational consequence still remain very elusive.

This paper is devoted to providing a new method for analyzing {\it fast-pairwise neutrino oscillation} \cite{2005PhRvD..72d5003S}. It is one of the collective neutrino oscillation modes and potentially generates strong and rapid flavor conversions. The linear analysis and some relevant numerical studies suggest that the electron neutrinos lepton number (ELN) angular crossings\footnote{In this paper, we only focus on the angular distributions of electron-type neutrinos ($\nu_e$) and their anti-partners ($\bar{\nu}_e$); equivalently, we assume that all heavy leptonic neutrinos have identical energy spectrum and angular distributions each other. They are in general, however, not true since the cross sections of neutrino-matter interactions depend on flavors, and the difference may be outstanding if on-shell muons appear in CCSN core \cite{2017PhRvL.119x2702B,2020PhRvD.102l3001F,2020PhRvD.102b3037G}. We refer readers to \cite{2020PhRvL.125y1801C,2021PhRvD.103f3013C} for possible influences of the heavy leptonic neutrinos in fast flavor conversion.} trigger the flavor conversion (see e.g., \cite{2017PhRvL.118b1101I,2017JCAP...02..019D,2021PhRvD.103f3002S,2021arXiv210101226B,2021arXiv210315267M,2021arXiv210312743S}). Thus, searching ELN-crossing in the neutrino data of theoretical CCSN models is the most straightforward way to judge whether the fast flavor conversion occurs in the CCSN environment.

The detailed angular information of each flavor of neutrinos is mandatory to carry out ELN-crossing search. However, high computational burden of CCSN simulations with multi-angle neutrino transport impedes the progress. Although there are several multi-D CCSN models with full Boltzmann neutrino transport at present \cite{2019PhRvD.100d3004A,2019PhRvD..99j3011D,2019ApJ...886..139N,2020PhRvD.101d3016A,2020PhRvD.101b3018D} (hereafter, the paper of \cite{2019ApJ...886..139N} is denoted as N19), they are nowhere near enough to scrutinize the progenitor- and time dependent features. More importantly, there are no available 3D radiation-hydrodynamic simulations with Boltzmann neutrino transport in the phase of interest for fast flavor conversions\footnote{A 3D CCSN simulation with full Boltzmann neutrino transport was performed in \cite{2020ApJ...903...82I}. However, it is up to the time of $\lesssim 20$ ms after bounce. In this phase, no fast-pairwise flavor conversions are expected due to strong suppression of $\bar{\nu}_e$ emissions (see \cite{2020PhRvR...2a2046M} for more details). We also note that the ELN search based on 3D full Boltzmann CCSN models in \cite{2019PhRvD.100d3004A,2020PhRvD.101d3016A} is not under fully consistent treatments. They solve neutrino transport on top of a fixed matter background, which is extracted from other CCSN simulations, until the field reaches steady state.}. On the other hand, one can run dozens of 3D simulations with covering the post-bounce phase up to $\sim 1$ s, if we use approximations in computation of neutrino transport. The most popular method is a two-moment approximation, in which the zero-th and first angular moments are solved with a closure relation for higher moments \cite{1972ApJ...171..127A,1981MNRAS.194..439T,2015MNRAS.453.3386J,2015PhRvD..91l4021F,2011PThPh.125.1255S,2019ApJS..241....7S,2021arXiv210202186L}. The angular degrees of freedom in neutrino momentum space are integrated out in this method, i.e., the number of dimensions in the neutrino transport is only four (three in space and one in neutrino energy). This approximation alleviates the computational cost substantially; consequently, two-moment methods have become nowadays standard in 3D CCSN models \cite{2016ApJ...831...98R,2016ApJS..222...20K,2018ApJ...865...81O,2018MNRAS.477L..80K,2019MNRAS.490.4622N,2019MNRAS.482..351V,2019ApJ...873...45G,2019PhRvC.100e5802S,2020MNRAS.491.2715B,2020MNRAS.492.5764N,2020MNRAS.494.4665P,2020MNRAS.498L.109M,2020arXiv201010506B}. It is, hence, worth to consider how we utilize CCSN models with two-moment neutrino transport for the analysis of fast flavor conversions.

In the last few years, great efforts have been devoted to developing surrogate methods to study fast flavor conversions based on neutrino data of two-moment neutrino transport. The {\it zero mode search} may be the simplest one. This method takes advantage of the following properties of the flavor conversion. The stability with respect to a homogeneous ($k=0$) mode in the corotating frame can be determined only from an inequality equation that is written as a function of the zero-th, first, and second angular moments \cite{2018PhRvD..98j3001D,2020PhRvD.101f3001G}, indicating that the simulations with two-moment neutrino transport provide sufficient information on the analysis. Another surrogate method was also proposed by \cite{2020JCAP...05..027A} (hereafter referred to as {\it polynomial method}). Similar as the zero mode search, it uses only a few angular moments of neutrinos. On the other hand, this method accesses to the angular structure of ELN by using a polynomial function of directional cosines of neutrino flight direction. In the method, ELN-crossings is identified by the sign of the angular integrated quantities (see \cite{2020JCAP...05..027A} for more details). This method was applied to neutrino data of 1D \cite{2021PhRvD.103f3013C} and 3D \cite{2021PhRvD.103f3033A} CCSN simulations, and the authors found positive signs of occurrence of ELN-crossings in these models.

Although these pioneer works pave the way to analyze fast flavor conversions in CCSN models  with two-moment neutrino transport, their accuracy and validity are another matter. Let us point out that both methods need to rely on closure relations, which can not be given accurately in the semi-transparent region of neutrinos by analytic prescriptions, however (see, e.g., \citep{2017MNRAS.469.1725M,2018ApJ...854..136N,2019ApJ...872..181H,2020ApJ...903...82I}). Next, we have found some cases that unstable modes appear at $k \neq 0$, whereas $k=0$ mode is stable (see, e.g., Fig.3 in N19), indicating that the zero mode search potentially leads misjudgement. For the polynomial method, on the other hand, the authors applied the method in the region with forward-peaked neutrino angular distributions. However, the higher-rank angular moments of neutrinos are completely neglected in the method. This would lead to misjudgements of ELN crossings, since the role of the high angular moments in characterizing the full angular distribution becomes more important with growing forward-peaked angular distributions.

Motivated by these concerns, we have recently scrutinized the capabilities of both methods \cite{2021arXiv210404106J,2021arXiv210405729N} (hereafter, the paper of \cite{2021arXiv210405729N} is denoted as N21). As predicted, we found that the angular distributions of incoming neutrinos are much less constrained by the zero-th and first angular moments than those of outgoing ones. The insensitiveness of incoming neutrinos to the low angular moments is an intrinsic limitation of moment methods; in other words it is a common issue among all ELN-crossing searches. For these reasons, we conclude that ELN-crossing searches with only using a few angular moments are not accurate, and they are only valid in the optically thick region.

This paper is devoted to get rid of the limitation; indeed, our new method presented in this paper is capable of detecting ELN-crossings in the region of forward-peaked angular distributions precisely. The essence of our idea is as follows. Let us first point out that the angular distributions of outgoing neutrinos can be reconstructed very well by using the method developed in N21. In other words, if we can overcome the shortage of reconstructing angular distributions of incoming neutrinos, the accuracy of ELN-crossing searches would be substantially improved. We tackle the issue by employing a ray-tracing method. One may wonder if the ray-tracing method is computationally expensive similar as full Boltzmann neutrino transport. Although it is true, we can substantially reduce the computational cost by following reasons. In previous studies, we have witnessed that the number of ELN-crossings in momentum space of neutrinos is usually one in CCSN environment (see, e.g., \cite{2019PhRvD.100d3004A} and N19).
In this case, the sign of ELN at $\mu = 1$ and $\mu = -1$, where $\mu$ denotes the radial directional cosines for neutrino flight directions with respect to the radial basis, is opposite each other. In other words, the appearance of ELN-crossings can be judged by checking the sign of ELN at $\mu = 1$ and $\mu = -1$. Since we can accurately reconstruct the ELN at $\mu =1$ by using our new method of N21, the ray-tracing neutrino transport along the direction of $\mu = -1$ provides a crucial information on judging ELN-crossings. It should also be mentioned that, even in (spatially) 3D CCSN models, the total number of rays along which we need to solve the transport equation is $N_{\theta} \times N_{\phi} \times N_{\varepsilon}$, where $N_{\theta}$, $N_{\phi}$, $N_{\varepsilon}$ denote the number of grid points of the lateral direction, azimuthal direction, and the neutrino energy, respectively. Since all rays can be solved independently (see Sec.~\ref{sec:method} for more details)\footnote{If we include neutrino-matter interactions of non-isoenergetic processes, the energy-coupling is inevitable in the ray-tracing method. However, the non-isoenergetic scatterings is subdominant in CCSN core; hence, we can safely neglect the contribution.}, our method is suitable for parallel computations. This also reduces the required computational time for ELN-crossing searches.

There is also another reason why the ray-tracing method is suited for our new method. As is well known, one of disadvantages of ray-tracing method is treatments of scatterings. To evaluate the inscattering contributions, the information on full angular distributions of neutrinos is mandatory in general, i.e., the transport equations along different rays are coupled. Since it requires iterative computations in general, the transport simulation becomes computationally expensive. In our method, however, we can overcome the difficulty by using the reconstructed angular distributions of neutrinos from the zero-th and first moments\footnote{As we shall discuss in Sec.~\ref{sec:method}, we consider isoenergetic scatterings in this paper. The reaction rate can be written in terms of zero-th and first angular moments. Hence, the reconstruction of full angular distributions is not always necessary to compute the inscattering contribution.}. This breaks coupling of transport equations and no iterative procedures are involved. This property is an important benefit in our hybrid approach.

This paper is organized as follows. We first describe the essence of our new method in Sec.~\ref{sec:method}. In Sec.~\ref{sec:demo}, we assess the capability (and validity) of our new ELN-crossing search by comparing results of 1D and 2D CCSN simulations with full Boltzmann neutrino transport. Finally, we summarize our conclusions in Sec.~\ref{sec:sumandconc}. We use the metric signature of $- + + +$. Unless otherwise stated, we work in units with $c=G=\hbar=1$, where $G$ and $\hbar$ denote the gravitational constant and the reduced Planck constant, respectively.

\section{Methods}\label{sec:method}
In our method, we start with reconstructing angular distributions of neutrinos from the zero-th and first angular moments obtained from CCSN simulations with two-moment neutrino transport. Although it is, in general, impossible to retrieve accurate angular distributions only from such lower angular moments, our previous study in N21 demonstrated a reasonable reconstruction. This success is based on the fact that there are some characteristic properties in the neutrino radiation field of CCSNe; indeed, some interesting correlations emerge between low angular moments and the full distributions in a neutrino data of CCSN model with Boltzmann neutrino transport \cite{2017ApJ...847..133R}. Taking advantage of the correlations, we determine some free parameters of a fitting function for which the shape of the angular distributions is determined solely from a flux factor ($\kappa$) (see N21 for more details)\footnote{The resultant fitting parameters are publicly available from the link: \url{https://www.astro.princeton.edu/~hirokin/scripts/data.html}}. The capability of our method was assessed by using a recent 2D CCSN model\footnote{In multi-D CCSN models, the angular distributions of neutrinos are no longer axisymmetric in momentum space. We, hence, employed the azimuthal-average angular distributions of neutrinos in the neutrino data of 2D CCSN simulations.} \cite{2019ApJ...880L..28N}. The demonstration illustrated the strength of our method; the angular distributions of outgoing neutrinos can be reconstructed accurately regardless of $\kappa$. In the present study, we take advantage of the merit; the ELN at $\mu = 1$ (along the outgoing radial direction) is computed from the reconstructed distribution of $\nu_e$ and $\bar{\nu}_e$ (see below for more details). We also underlined a weakness of the method; it is not capable of determining angular distributions of incoming neutrinos for large $\kappa$, i.e., forward-peaked distributions. In this paper, we use a ray-trace method to compensate for the shortage. We refer readers to N21 for more details of our reconstructed method.

In the present method, we judge appearances of ELN-crossing by comparing the sign of ELN at $\mu=1$ (outgoing) and $\mu=-1$ (incoming) angular points. The ray-tracing transport is in charge of computing the ELN at $\mu=-1$. If the sign of the ELNs is opposite each other, it indicates the appearance of ELN-crossing. On the other hand, if the sign is the same, we can not make a robust judgement in general, since this corresponds to the case either no crossings or the even number of crossings. This shortage can be improved by solving radiation transport equations for different angles, i.e., we need to increase the number of rays. However, the computational cost would be increased accordingly, which reduces the merit (cheap computation) of our new method. It should be pointed out that recent ELN-crossing searches based on CCSN models with full Boltzmann neutrino transport suggest that the number of ELN-crossings is usually one in CCSN core (see, e.g., N19). 
Therefore, we judge no ELN crossings if the sign is the same in this study, although we need to keep in mind the uncertainty.

In our ray-tracing computations, we impose several approximations; the space-time is flat; no fluid-velocity dependences are taken into account; neutrino radiation fields have already settled into a steady-state. Under these assumptions, the transport equation can be written as an ordinary differential equation,
\begin{equation}
- \frac{d}{dr} f_{\rm in} (\varepsilon,r) = \Bigl( \frac{\delta f_{\rm in}}{\delta t} \Bigr)_{\rm col} (\varepsilon,r),
\label{eq:basiceq_raytrace}
\end{equation}
where $\varepsilon$, $r$, and $t$ denote the energy of neutrinos, radius, and time, respectively. $f_{\rm in}$ represents the distribution function of neutrinos ($f$) in the direction of $\mu=-1$. The right hand side of Eq.~\ref{eq:basiceq_raytrace} represents the neutrino-matter interactions. If the reaction rates employed in CCSN simulations with two-moment neutrino transport are available in the output data, they may be directly used for this ray-tracing computation. Otherwise, we need to take the fluid data from the output, and then compute each reaction rate by a post-processing manner. Although it is commended to employ the same weak interaction rates used in CCSN simulations, this may result in increasing the computational cost. For the usability purpose, we recommend to use a minimum but essential set of weak interactions in the post-shock region of CCSN core: electron captures by free protons (and the inverse reaction), positron captures by free neutrons (and the inverse reaction), scatterings to nucleons, and coherent ones to heavy nuclei\footnote{We compute these reaction rates following by \cite{1985ApJS...58..771B}.}. One may wonder if thermal processes such as electron-positron pairs and nucleon-nucleon bremsstrahlung processes need to be taken into account. However, those reactions are important only in optically thick region where the angular distributions of neutrinos (including incoming directions) can be well reconstructed (see N21). This indicates that our ray-tracing method is not necessary in this region; hence we stop the ray-tracing computation before entering the optically thick region (see below for more details). Therefore, the ignorance of the thermal processes in the ray-tracing method is a reasonable approximation. Indeed, we will show that ELN-crossing searches under the approximations lead to consistent judgement to the result of full Boltzmann neutrino transport.

We remark on the computation of scatterings in our method, since it generally requires a special attention for ray-tracing transport. The collision term of scatterings in Eq.~\ref{eq:basiceq_raytrace} can be written as a sum of inscattering and outscattering components. They can be written as
\begin{equation}
\begin{aligned}
& \Bigl( \frac{\delta f_{\rm in}}{\delta t} \Bigr)_{\rm inscat} = \frac{(\varepsilon)^{2}}{(2 \pi)^3} \int d \Omega^{\prime} R_{\rm scat} (\Omega^{\prime}) f (\Omega^{\prime}), \\
& \Bigl( \frac{\delta f_{\rm in}}{\delta t} \Bigr)_{\rm outscat} = - \frac{(\varepsilon)^{2}}{(2 \pi)^3} f_{\rm in} \int d \Omega^{\prime} R_{\rm scat} (\Omega^{\prime}),
\end{aligned}
\label{eq:isoenergyscatkenel}
\end{equation}
where $R_{\rm scat}$ denotes the scattering kernel (which can be computed by fluid data). We refer readers to \cite{2014ApJS..214...16N} for more general expressions. As shown in Eq.~\ref{eq:isoenergyscatkenel}, the out-scattering component can be computed without any problems, since the required information on neutrino distribution function is only $f_{\rm in}$, which is obtained from the transport equation along the same radial ray. On the other hand, the inscattering component depends on $f$ on different angular directions, indicating that the transport equation can not be closed by the single ray. In this study, we employ $f$ reconstructed from the method of N21 from the zero-th and first angular moments in the computation of inscattering rates\footnote{We note that the angular dependence of the scattering kernel, $R_{\rm scat}$, for both nucleon and heavy nuclei is up to the first order of directional cosine of the scattering angle unless the energy exchange is taken into account, indicating that the angular integration with $f$ can be written in terms of zero-th and first angular moments. Hence, it is also possible to evaluate the inscattering component by directly using the moment data extracted from CCSN simulations.}.

\begin{figure*}
    \includegraphics[width=\linewidth]{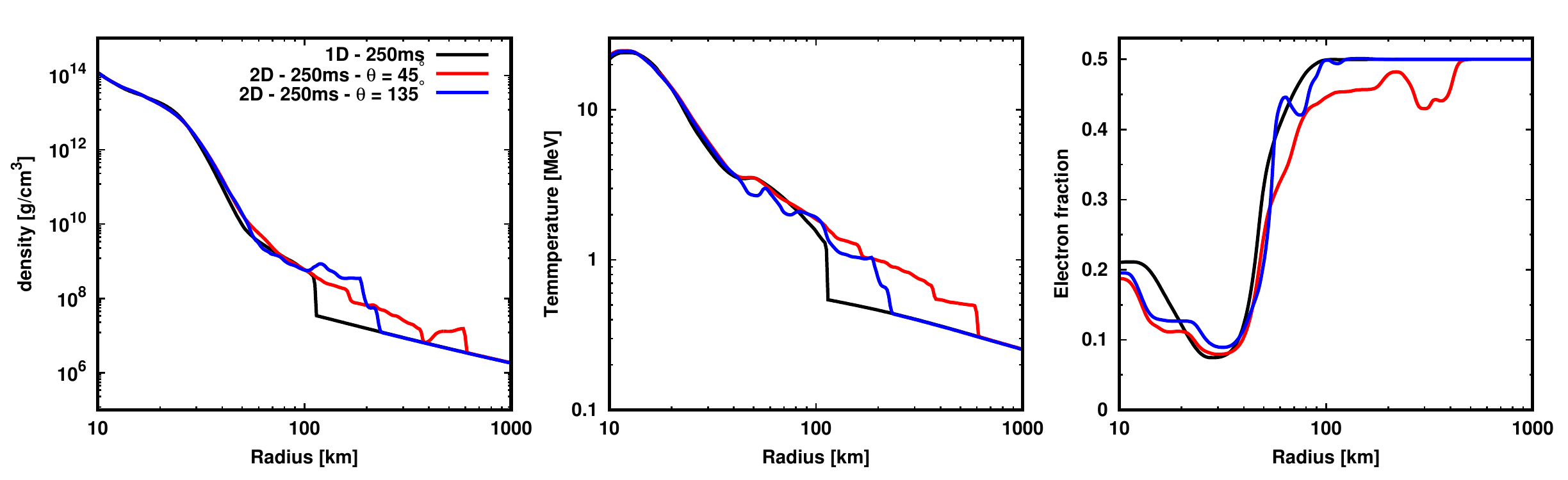}
    \caption{From let to right panels, we show the radial profiles of baryon mass density, matter temperature, and electron fraction, respectively, for CCSN models employed in this study. The detailed analyses can be seen in \cite{2019ApJS..240...38N} and \cite{2019ApJ...880L..28N} for 1D and 2D model, respectively. In this paper, we selected the two radial rays ($\theta = 45^{\circ}$ and $135^{\circ}$) in the 2D model as representative examples.}
    \label{graph_radialprofile_hydro}
\end{figure*}

\begin{figure*}
    \includegraphics[width=\linewidth]{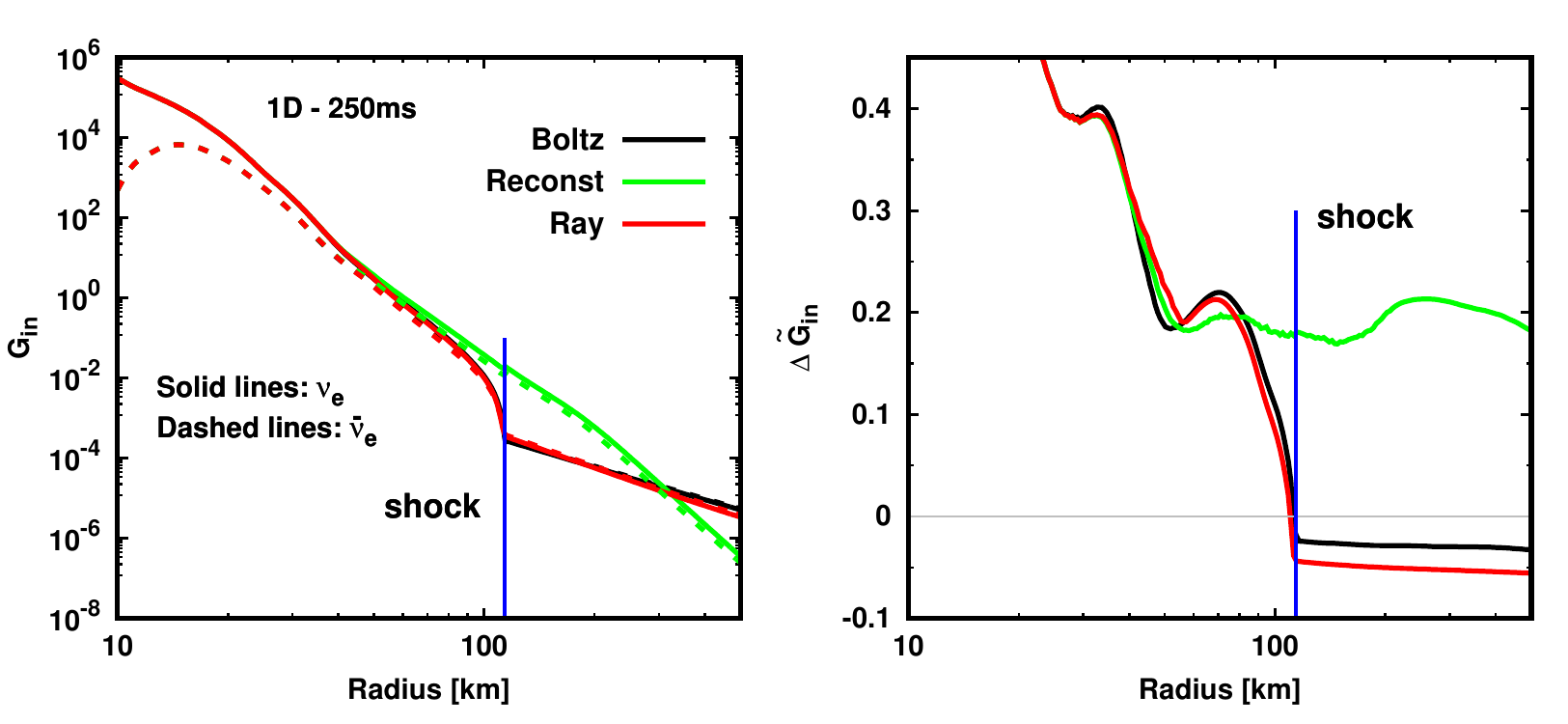}
    \caption{Left: radial profile of $G_{\rm in}$ for 1D CCSN model. Colors distinguish the difference in computing $G_{\rm in}$. The black line represents the result of $G_{\rm in}$ computed from $f$ given by CCSN simulations with full Boltzmann neutrino transport. For the light-green one, $G_{\rm in}$ is computed from the reconstructed $f$ from the zero-th and first moments by using the method of N21. The red one corresponds to the result computed by our new method presented in this paper. The line type distinguishes the neutrino species; solid and dashed lines are for $\nu_e$ and $\bar{\nu}_e$, respectively. 
Right: the same as the left panel but for $\Delta \tilde{G}_{\rm in}$. It is $\Delta G_{\rm in}$ normalized by the summation of $G_{\rm in}$ over $\nu_e$ and $\bar{\nu}_e$.  As a reference, the shock radius is displayed with a blue line in both panels. We also highlight $\Delta \tilde{G}_{\rm in}=0$ by a gray line in the plot.
}
    \label{graph_radialprofile_Gincome_1D}
\end{figure*}

\begin{figure}
    \includegraphics[width=\linewidth]{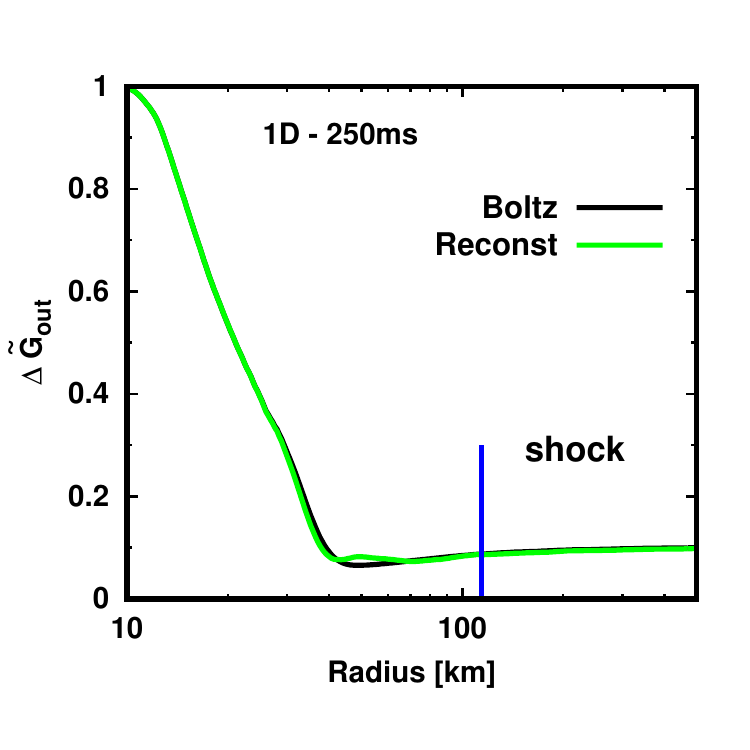}
    \caption{Same as the right panel of Fig.~\ref{graph_radialprofile_Gincome_1D} but for $G_{\rm out}$. We note that our new method adopts $f_{\rm out}$ reconstructed from the zeroth and first angular moments in the computation of $G_{\rm out}$; hence we omit the red line in this figure. $\Delta \tilde{G}_{\rm out}$ denotes $\Delta G_{\rm out}$ normalized by the sum of $G_{\rm out}$ over $\nu_e$ and $\bar{\nu}_e$.
}
    \label{graph_radialprofile_deltaGout_1D}
\end{figure}

\begin{figure}
    \includegraphics[width=\linewidth]{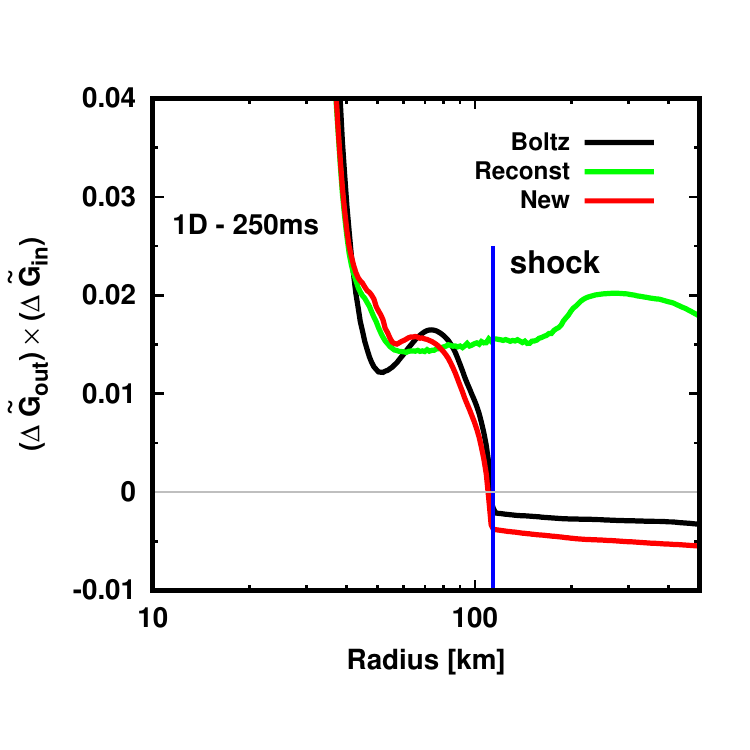}
    \caption{Radial profile of $\Delta \tilde{G}_{\rm in} \times \Delta \tilde{G}_{\rm out}$ for 1D model. The color coding is the same as Fig.~\ref{graph_radialprofile_Gincome_1D}. We note that the title of red line is replaced from ``Ray'' to ``New'', since the previous title is misleading in this plot ($\tilde{G}_{\rm out}$ is computed from the reconstructed method of N21). See text for more detail.
 }
    \label{graph_radialprofile_ELNcross_1D}
\end{figure}

We solve Eq.~\ref{eq:basiceq_raytrace} from the outer boundary of CCSN simulation to the radially inward direction. As we have already pointed out, the ray-tracing neutrino transport is necessary only in optically thin region where neutrino angular distributions are forward-peaked. We, thus, connect the obtained $f_{\rm in}$ to that reconstructed from zeroth and first moments; the detail of the procedure is as follows. We first prepare two threshold flux factors: $\kappa_1$ and $\kappa_2$ ($\kappa_1 < \kappa_2$). In the region of $\kappa < \kappa_1$ (optically thick region), $f_{\rm in}$ is determined from the reconstructed distribution ($f_{\rm in}^{\rm reco}$). In the region of $\kappa > \kappa_2$, we employ the solution of ray-tracing neutrino transport ($f_{\rm in}^{\rm RT} $). In the intermediate region ($\kappa_1 < \kappa < \kappa_2$), we determine $f_{\rm in}$ by mixing the two solutions. More specifically, it is obtained as,
\begin{equation}
f_{\rm in} = q(\kappa) f_{\rm in}^{\rm RT} + \Bigl( 1-q(\kappa) \Bigr) f_{\rm in}^{\rm reco} (\kappa),
\label{eq:fin_interpo}
\end{equation}
where
\begin{equation}
q(\kappa) = \frac{\kappa - \kappa_1}{\kappa_2 - \kappa_1}.
\label{eq:qdef}
\end{equation}
As shown above, the ray-tracing neutrino transport is necessary in the region of $\kappa > \kappa_1$.

The two parameters, $\kappa_1$ and $\kappa_2$, are determined following the result of N21. We found that $f_{\rm in}^{\rm reco}$ agrees reasonably well to results of Boltzmann neutrino transport at $\kappa \lesssim 0.4$, whereas it deviates from the original at $\kappa \gtrsim 0.5$; hence, we adopt $\kappa_1 = 0.4$ and $\kappa_2 = 0.5$ in this study. Since the neutrino-matter interactions sensitively depend on neutrino energy, the matching region should also be varied with neutrino energy. Our method is capable of capturing such an energy-dependent feature (as long as the neutrino data provided by CCSN simulations is energy dependent.). Although ELN-crossing searches are made in the energy-integrated quantities, the energy-dependent treatment is beneficial to draw the angular structure of neutrinos accurately.

\section{Demonstrations}\label{sec:demo}
In this section, we discuss the capability of our new method by demonstrating ELN-crossing searches by using neutrino data of 1D and 2D CCSN simulations with full Boltzmann neutrino transport. We employ most recent CCSN models of a $11.2$ solar mass progenitor at a time snapshot of 250 ms after bounce. The detailed features of CCSN dynamics were discussed in our previous papers: 1D \cite{2019ApJS..240...38N} and 2D \cite{2019ApJ...880L..28N}. In the raw neutrino data, we found no ELN-crossings in the post-shock region for the 1D model. On the other hand, we found the crossings in the 2D data (we refer readers to N19 for the detailed analyses).

\begin{figure*}
    \includegraphics[width=\linewidth]{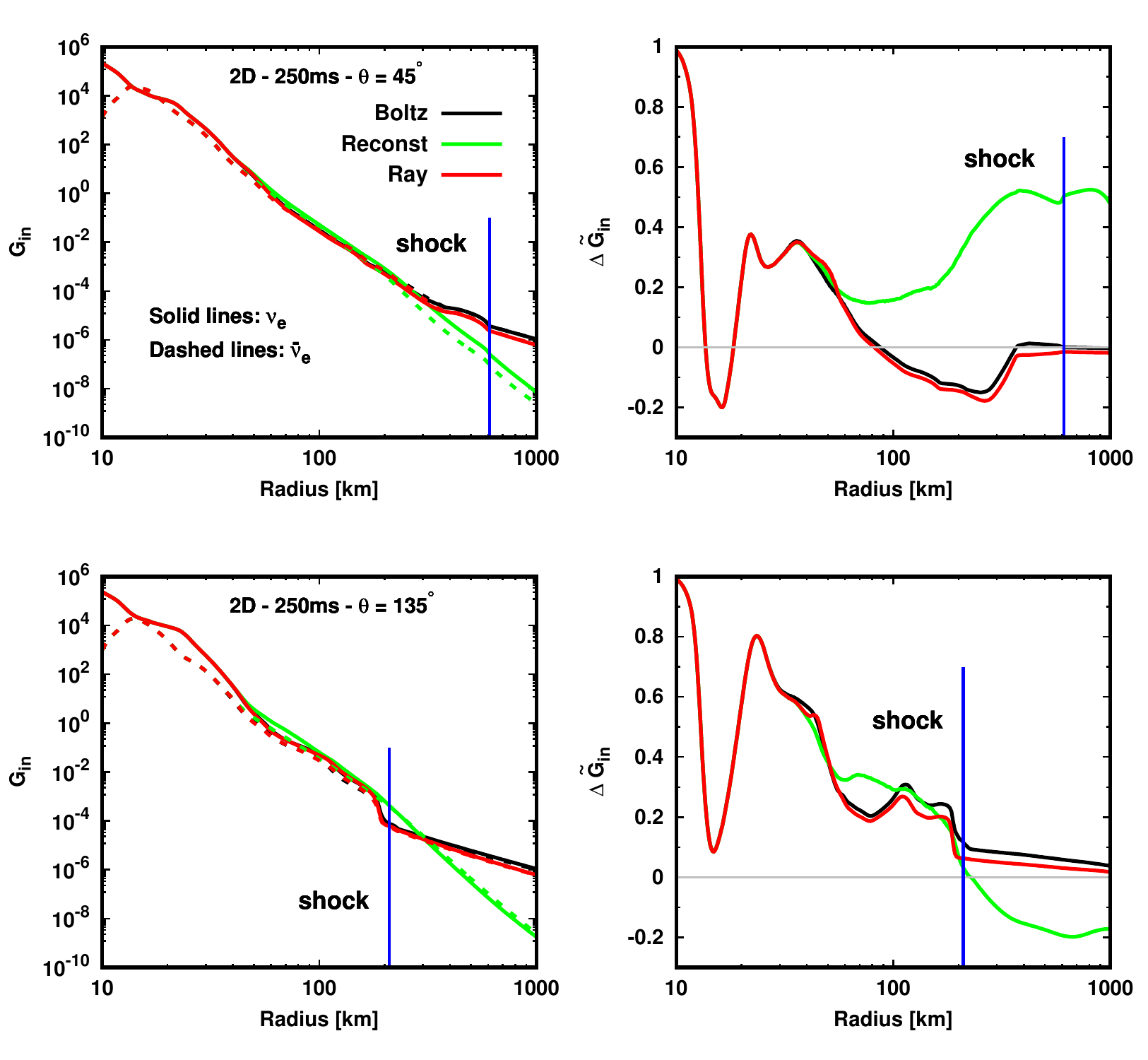}
    \caption{Same as Fig.~\ref{graph_radialprofile_Gincome_1D} but for 2D CCSN model. Top and bottom panels show the result along the radial ray of $\theta = 45^{\circ}$ and $135^{\circ}$, respectively.
 }
    \label{graph_radialprofile_Gincome_2D}
\end{figure*}

\begin{figure*}
    \includegraphics[width=\linewidth]{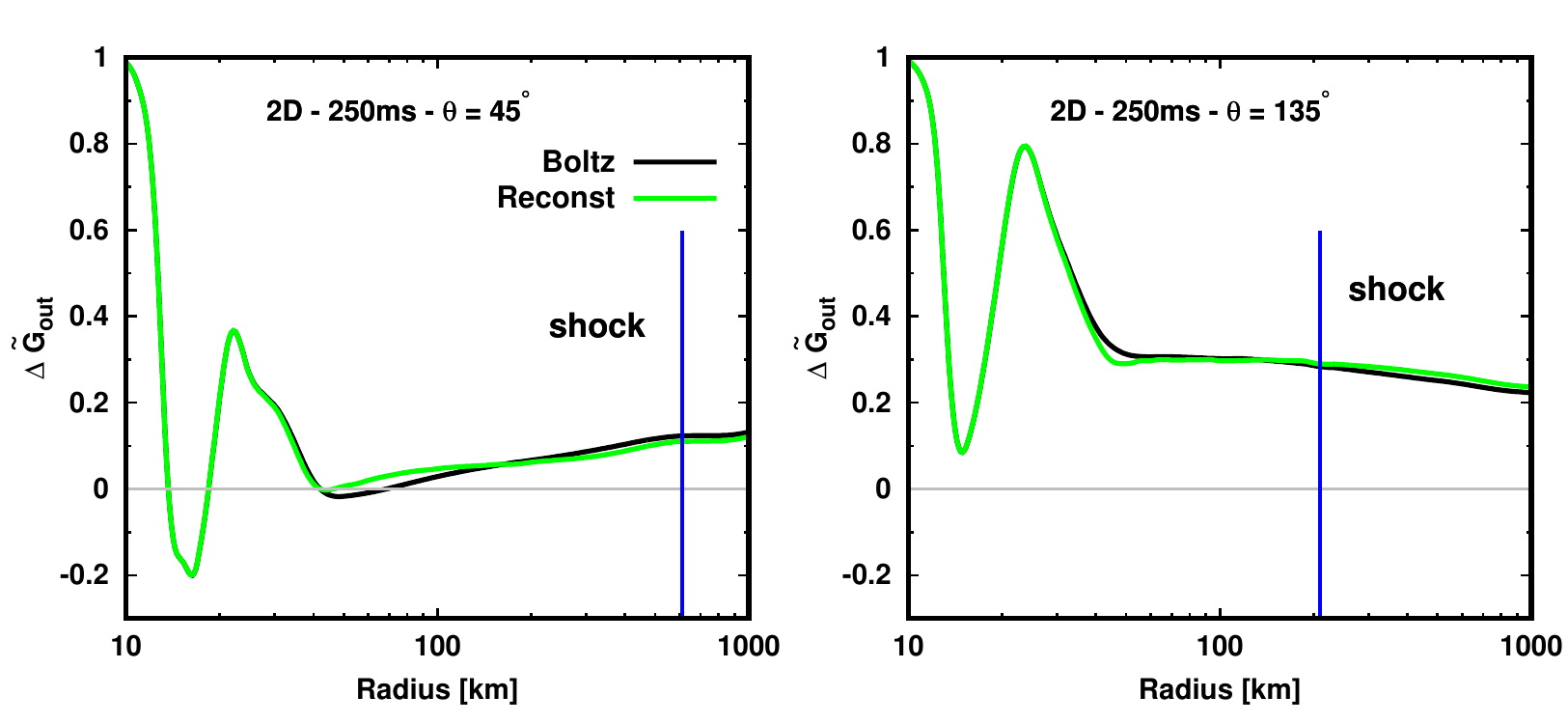}
    \caption{Same as Fig.~\ref{graph_radialprofile_deltaGout_1D} but for 2D CCSN model. Left and right panels display the result along the radial ray with $\theta = 45^{\circ}$ and $135^{\circ}$, respectively.
 }
    \label{graph_radialprofile_deltaGout_2D}
\end{figure*}

\begin{figure*}
    \includegraphics[width=\linewidth]{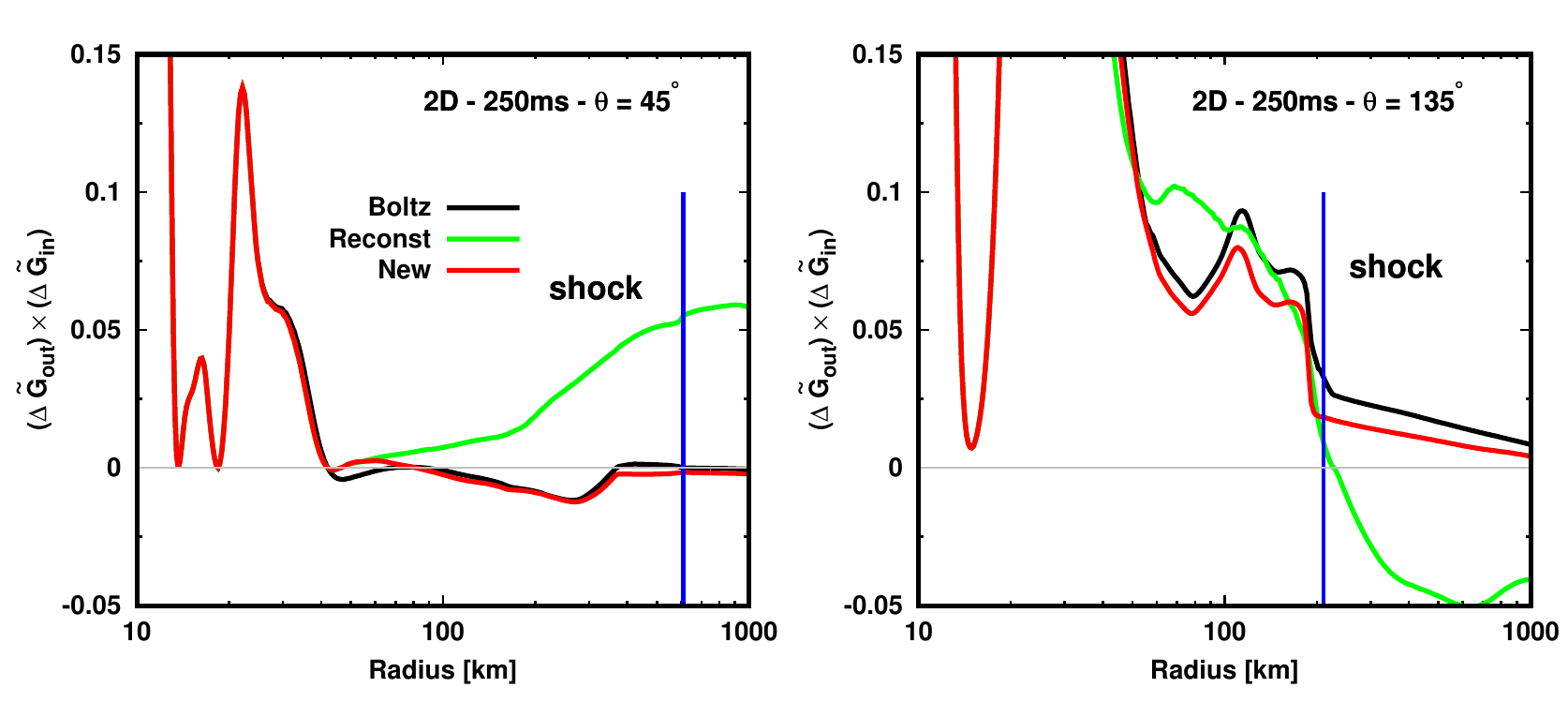}
    \caption{Same as Fig.~\ref{graph_radialprofile_ELNcross_1D} but for 2D CCSN model.
 Left and right panels display the result along the radial ray with $\theta = 45^{\circ}$ and $135^{\circ}$, respectively.
}
    \label{graph_radialprofile_ELNcross_2D}
\end{figure*}

We first compute the energy-dependent zero-th and first angular moments of neutrinos from the distribution function of neutrinos ($f$) obtained from these simulations. We treat these moments as the input data for our method. The fluid data is also extracted from the simulations (see Fig.~\ref{graph_radialprofile_hydro}) to compute the reaction rates of neutrino-matter interactions (see Sec.~\ref{sec:method}). In the following analyses, we use $G_{\rm in}$ and $G_{\rm out}$ to quantify the capability of our ELN-crossing search, which are defined as
\begin{equation}
\begin{aligned}
&G_{\rm in} = \int d (\frac{\varepsilon^3}{3}) f_{\rm in}(\varepsilon), \\
&G_{\rm out} = \int d (\frac{\varepsilon^3}{3}) f_{\rm out}(\varepsilon), \\
\end{aligned}
\label{eq:def_GinGout}
\end{equation}
where $f_{\rm out}$ denotes the distribution function $f$ in the direction of $\mu=1$. The energy-integration in the right hand side of Eq.~\ref{eq:def_GinGout} is carried out with the unit of MeV. We also compute the difference of $G$ between $\nu_e$ and $\bar{\nu}_e$,
\begin{equation}
\Delta G = G_{\nu_e} - G_{\bar{\nu}_e},
\label{eq:def_deltaG}
\end{equation}
while we omit the index of "${\rm in}$" and "${\rm out}$" in the expression. $\Delta G_{\rm in}$ and $\Delta G_{\rm out}$ are the most important variables in our method, since they are directly associated with ELN crossings. The former and latter represents the dominance of $\nu_e$ relative to $\bar{\nu}_e$ at $\mu=-1$ (incoming) and $\mu=1$ (outgoing) directions, respectively. In the case with the sign of the two variables is opposite, i.e., $\Delta G_{\rm out} \Delta G_{\rm in} < 0$, $\nu_e$ is dominant over $\bar{\nu}_e$ in either $\mu=-1$ or $1$ direction, meanwhile $\bar{\nu}_e$ overwhelms $\nu_e$ in the other direction, suggesting that ELN crossings occur in this case (see Sec.~\ref{sec:method} for more details).

Figure~\ref{graph_radialprofile_Gincome_1D} shows the radial profile of $G_{\rm in}$ (left panel) and $\Delta \tilde{G}_{\rm in}$ (right panel) for 1D CCSN model. $\Delta \tilde{G}_{\rm in}$ represents $\Delta G_{\rm in}$ normalized by the sum of $G_{\rm in}$ over $\nu_e$ and $\bar{\nu}_e$ at the same radius. This figure illustrates how much the ray-tracing treatment improves the computation of $G_{\rm in}$. In the left panel, $G_{\rm in}$ computed by the new method (red line) is almost identical to the original (black line), whereas the deviation is conspicuous at $\gtrsim 50$ km for one computed from the reconstructed $f_{\rm in}$ from the zeroth and first moments (light-green line).

As shown in the left panel of Fig.~\ref{graph_radialprofile_Gincome_1D}, the difference of $G_{\rm in}$ between $\nu_e$ and $\bar{\nu}_e$ is subtle at $\gtrsim 50$ km. This fact indicates that the high-fidelity computation of $G_{\rm in}$ is required to reproduce the sign of $\Delta G_{\rm in}$ accurately, which is more clearly displayed in the right panel of Fig.~\ref{graph_radialprofile_Gincome_1D}. The difference of $G_{\rm in}$ between $\nu_e$ and $\bar{\nu}_e$ is a few percent of their summation (see black line in the right panel). Nevertheless, the new method reproduces the result of the original (compare red and black lines). On the other hand, the reconstructed counterpart (light-green line) generates qualitatively different $\Delta G_{\rm in}$; indeed, the sign of $\Delta G_{\rm in}$ becomes opposite from the original at the pre-shock region.

In Fig.~\ref{graph_radialprofile_deltaGout_1D}, we compare the radial profile of $\Delta \tilde{G}_{\rm out}$ obtained by our new method to that of the original. Similar to $\Delta \tilde{G}_{\rm in}$, $\Delta \tilde{G}_{\rm out}$ denotes $\Delta G_{\rm out}$ normalized by the sum of $G_{\rm out}$ over $\nu_e$ and $\bar{\nu}_e$ at the same radius. In the computation of $G_{\rm out}$, we employ the reconstructed angular distribution from the zero-th and first angular moments by following the method of N21. As mentioned already, the angular distributions of outgoing neutrinos can be well reconstructed without any extra prescriptions (see Sec.~\ref{sec:intro}). As expected, we confirm that  $\Delta G_{\rm out}$ computed from the reconstructed distributions is very similar as that of the original, which are displayed in Fig.~\ref{graph_radialprofile_deltaGout_1D}.

Both $\Delta G_{\rm in}$ and $\Delta G_{\rm out}$ computed by our new method show a good agreement with the originals, indicating that 
their product, i.e., $\Delta G_{\rm out} \times \Delta G_{\rm in}$
is also captured accurately, which is displayed in Fig.~\ref{graph_radialprofile_ELNcross_1D}. In the original profile (black line), the sign of the product is positive in the post-shock region, indicating that either $\nu_e$ or $\bar{\nu}_e$ is dominant in the entire neutrino flight direction\footnote{In this model, $\nu_e$ is dominant in the region.}, i.e., there are no-ELN crossings in the region. On the contrary, the sign of the product is negative in the pre-shock region; thus, the dominance of $\nu_e$ or $\bar{\nu}_e$ depends on the neutrino angle, indicating that ELN-crossings appear. The key player to generate ELN-crossings at pre-shock regions is a coherent scatterings of heavy nuclei, the detail of the physical mechanism is described in our previous paper \cite{2020PhRvR...2a2046M}. Our new method detects the ELN-crossing precisely, whereas one without ray-tracing method does not. It should be emphasized that $\Delta G_{\rm out}$ is common for both methods, indicating that $\Delta G_{\rm in}$ is responsible for the difference. This is a strong evidence that our ray-tracing method qualitatively improves the accuracy of ELN-crossing searches.

We now turn our attention to the 2D model. It should be stressed that the spatial dimension does not change the procedure of our method, since we solve transport equations independently along each radial ray toward inward direction ($\mu=-1$). In the following analysis, we focus on two radial rays with $\theta = 45^{\circ}$ and $135^{\circ}$ in 2D space. We note that ELN crossings appear in the radia ray with $\theta = 45^{\circ}$ in the original neutrino data with Boltzmann transport, whereas there are no ELNCrossing at $\lesssim 1000$ km in the ray with $\theta = 135^{\circ}$ (see also Fig.2 in N19). By applying our new method to search for ELN crossings along the two radial rays, we can assess the capability of our method in both cases with and without ELN-crossings.

Figure~\ref{graph_radialprofile_Gincome_2D} portrays the radial distribution of $G_{\rm in}$ and $\Delta \tilde{G}_{\rm in}$ computed based on our new method (red line), reconstructed $f$ from the zero-th and first moments (light-green line), and those given by Boltzmann neutrino transport (black line), the trend of which is essentially the same as that reported in our 1D model (see Fig.~\ref{graph_radialprofile_Gincome_1D}). We confirm that the new method substantially improves the computations of $G_{\rm in}$. We have to remark a caveat, however. At the radius of $\sim 500$ km along the radial ray with $\theta = 45^{\circ}$, our new method shows a different sign of $\Delta G_{\rm in}$ from that in the original (see top and right panel of Fig.~\ref{graph_radialprofile_Gincome_2D}). This is attributed to the fact that the difference of $G_{\rm in}$ between $\nu_e$ and $\bar{\nu}_e$ is so tiny. As a result, a small error in our method leads to misjudgements of the sign. This error may be reduced if we increase the level of approximations in our method. However, it would lead to increase the computational burden, which is undesirable for approximate ELN-crossing searches. To make more reliable judgement for such delicate ELN-crossings, more complete treatments of neutrino transport, i.e., full Boltzmann neutrino transport would be indispensable.

In Fig.~\ref{graph_radialprofile_deltaGout_2D}, we show the result of $\Delta \tilde{G}_{\rm out}$ as a function of radius. Let us emphasize again that $\Delta G_{\rm out}$ is obtained from the reconstructed $f$ from the zero-th and first angular moments. This picture represents the good capability of our method. On the other hand, we again find a failure of capturing the sign of $\Delta G_{\rm out}$, at $\sim 50$ km in the radial ray of $\theta = 45^{\circ}$ (left panel of Fig.~\ref{graph_radialprofile_deltaGout_2D}). Similar as the above argument, this error is attributed to the fact that the $G_{\rm out}$ of $\nu_e$ and $\bar{\nu}_e$ is almost identical. Nevertheless, in most of the spatial regions, the basic feature of $\Delta G_{\rm out}$ can be well captured by the new method.

The radial profile of $\Delta \tilde{G}_{\rm out} \times \Delta \tilde{G}_{\rm in}$ in the 2D model is displayed in Fig.~\ref{graph_radialprofile_ELNcross_2D}. This illustrates that our new method provides the same judgement of ELN-crossings as that made by Boltzmann neutrino transport along a radial ray with $\theta = 135^{\circ}$. This is based on the fact that $\nu_e$ number flux is substantially larger than that of $\bar{\nu}_e$ along the radial ray, which is due to the coherent asymmetric neutrino emissions (see \cite{2019ApJ...880L..28N} for more details). As a result, there are no delicate competitions between $\nu_e$ and $\bar{\nu}_e$ for both inward- and outward- neutrino flight directions; hence, our method is capable of providing a robust diagnostics of ELN-crossings. It should be stressed, however, that the judgment is failed if we do not use a ray-tracing method. The low accuracy of reconstructing incoming neutrinos ($G_{\rm in}$) is mainly responsible for the misjudgement (see bottom panels of Fig.~\ref{graph_radialprofile_Gincome_2D}). In the radial ray with $\theta = 45^{\circ}$, on the other hand, we find misjudgement even in the new method at $\sim 50$ km and $\sim 500$ km. For the former and latter, it is due to the error of $G_{\rm out}$ and $G_{\rm in}$ in our method, respectively; the reason of which was already discussed. This result suggests that our new method is capable of detecting ELN-crossings unless the crossing is so subtle (a few percents) in reality. We have to keep in mind the uncertainty and limitation when we apply our method for ELN-crossing searches.

\section{Summary and conclusion}\label{sec:sumandconc}
There is growing evidence that the appearance of ELN-crossings in angular distributions of neutrinos are precursors of occurring fast-pairwise neutrino oscillation. This indication heightens the awareness of the importance of multi-angle treatments of neutrino transport. However, the available CCSN models with full Boltzmann neutrino transport are still limited by their high computational burdens, which have inhibited the progress of the detailed analysis. On the other hand, there are many multi-d CCSN models with approximate neutrino transport, which covers the long-term post-bounce phase for various types of CCSN progenitors. This has motivated the community to develop surrogate methods to determine the occurrence of fast flavor conversions under limited information on neutrino radiation fields. The neutrino data is usually a few rank of angular moments; hence it is interesting to consider how we can utilize them in the analysis of flavor conversions.

There are some previous works tackling the issue. It turned out very recently, however, that they are not capable of identifying the occurrence of fast flavor conversions in the region of forward-peaked angular distributions. The source of the problem is that the low angular moments are insensitive to the incoming neutrinos in such regions. In this paper, we propose to use a ray-tracing method to compensate for the shortage. In this method, we determine the ELN at $\mu=1$ (outgoing neutrinos) by the reconstructed angular distributions of neutrinos from the zero-th and first moments by using the method of N21. On the other hand, the determination of ELN at $\mu=-1$ is complemented by the ray-tracing method. In Sec.~\ref{sec:demo}, we assess the capability of our method by making a detailed comparison to results of full Boltzmann neutrino transport in 1D and 2D. We demonstrate that our new method improves the accuracy of ELN-crossing searches substantially. Unless the crossing is very subtle (a few percent of the sum of the occupation number of $\nu_e$ and $\bar{\nu}_e$), our method is capable of making accurate judgement of ELN-crossings.

We next apply our new method presented in this paper to multi-D CCSN models with multi-group two moment neutrino transport. The results will be discussed in a separate paper.

\section{Acknowledgments}
We acknowledge conversations with Sherwood Richers, Sam Flynn, Nicole Ford, Evan Grohs, Jim Kneller, Gail McLaughlin, Don Willcox, Taiki Morinaga, Eirik Endeve, and Adam Burrows. H.N acknowledges support from the U.S. Department of Energy Office of Science and the Office of Advanced Scientific Computing Research via the Scientific Discovery through Advanced Computing (SciDAC4) program and Grant DE-SC0018297 (subaward 00009650). The numerical computations of our CCSN models were performed on the K computer at the RIKEN under HPCI Strategic Program of Japanese MEXT (Project ID: hpci 160071, 160211, 170230, 170031, 170304, hp180179, hp180111, and hp180239). L.J. acknowledges support provided by NASA through the NASA Hubble Fellowship grant number HST-HF2-51461.001-A awarded by the Space Telescope Science Institute, which is operated by the Association of Universities for Research in Astronomy, Incorporated, under NASA contract NAS5-26555.

\bibliography{bibfile}


\end{document}